\documentstyle[pra,aps,preprint,12pt]{revtex}
\tightenlines
\newcommand{\halb}{\mbox{\large${1\over2}$}}

\newcommand{\subsub}[3]{{#1_{\hskip-1pt\scriptscriptstyle #2}}_{{ }_{\hskip-0pt \scriptscriptstyle #3}}\hskip0pt}

\newcommand{\fett}[1]{\mbox{\boldmath $#1$}}
\newcommand{\Fett}[1]{\mbox{\boldmath\tiny $#1$}}

\begin{document} 
\draft 
\title
{One-Nucleon Effective Generators of the Poincar\'e Group derived from a Field Theory: Mass Renormalization}
\author
{
A. Kr\"uger and W. Gl\"ockle\\
{\it Institut f\"ur Theoretische Physik II, Ruhr-Universit\"at Bochum,\\
D-44780 Bochum, Germany}
}
\maketitle

\begin{abstract}
We start from a Lagrangian describing scalar "nucleons" and mesons
which interact through a simple vertex.
Okubo's method of unitary transformation is used to describe a single
nucleon dressed by its meson cloud. We find an expression for the
physical mass of the nucleon being correct up to second order in the
coupling constant. It is then verified that this result is the same as
the corresponding expression found by Feynman techniques. Finally we
also express the three boost operators in terms of the physical
nucleon mass. Doing so we find expressions for all the ten generators
of Poincar\'e transformations for the system of one single dressed
nucleon.
\end{abstract}

\section{Introduction}
In a previous study \cite{krgl} we provided formal expressions for  N-nucleon
effective generators of the Poincar\'e group derived from a field theory.
The ten hermitian generators derived from standard hermitian Lagrangians
describing interacting fields were blockdiagonalized by one and the same
unitary transformation, such that the space of a fixed number of nucleons
is separated from the rest of the space. The existence proof we carried
through makes use of 
a formal power series expansion in the coupling constant g.
In this article we want to apply this procedure to the case of a
one-nucleon subspace. We shall study the effective generators in that
space in the first nontrivial order ${\cal O}(g^2)$. As
has to be expected this implies the question of mass renormalization in
the Hamiltonian and the boost operators, which are the four generators
carrying interactions in the instant form of relativistic dynamics.

The question will be, how the mass renormalization will come about and
whether after that step the Poincar\'e algebra for those effective
generators will still be valid.

We investigate these problems with the help of a Lagrangian describing
interacting scalar "nucleons" and mesons. In Section \ref{conmassren}
we formulate
the steps taken within the unitary transformation leading to the
effective generators. The mass renormalization in the effective
one-nucleon Hamiltonian is carried through in Section
\ref{massrenormalisation}. 
An important
but lengthy algebra demonstrating  the momentum independence of the mass
shift is deferred to the Appendix. Our result gained in the Hamiltonian
formalism is identical to the one gained by standard Feynman
methods in a manifestly covariant manner. This is shown in Section
\ref{compfeyn}. 
The renormalisation of the effective boost operators is carried
through in Section \ref{renbo}. Finally we summarize in section \ref{summary}.

\section{Conditions for mass renormalisation}\label{conmassren}

Poincar\'e invariance requires that there exists a unitary representation of
the Poincar\'e group defined in a Hilbert space. The corresponding ten
generators fulfill the following set of commutation relations
\begin{eqnarray}
&&[{ P}_{i},H]=0\label{lie1}\\
&&[{ J}_{i},H]=0\label{lie2}\\
&&[{ P}_{i},{ P}_{j}]=0\label{lie3}\\
&&[{ J}_{i},{ J}_{j}]=i\epsilon_{ijk}{ J}_{k}\label{lie4}\\
&&[{ J}_{i},{ P}_{j}]=i\epsilon_{ijk}{ P}_{k}\label{lie5}\\
&&[{ J}_{i},{ K}_{j}]=i\epsilon_{ijk}{ K}_{k}\label{lie6}\\
&&[H,{ K}_{i}]=-i { P}_{i}\label{lie7}\\
&&[{ K}_{i},{ K}_{j}]=-i\epsilon_{ijk}{ J}_{k}\label{lie8}\\
&&[{ P}_{i},{ K}_{j}]=-i\delta_{ij}H\label{lie9}
\end{eqnarray}
Here $H$ is the Hamilton operator, $P_i$, and $J_i$ the three components
of the momentum and angular momentum operators and $K_i$ the three
boost operators. We note that in the instant form the four operators
$H$ and $K_i$ carry interactions. From Eqs. (\ref{lie1}) and (\ref{lie3}) 
we see that $H$ and $P_i$ can have simultaneous eigenstates which we
denote as $|\Psi_{\Fett p}\rangle$:
\begin{eqnarray}
H|\Psi_{\Fett p}\rangle&=&E|\Psi_{\Fett p}\rangle\label{heigen}\\
P_i|\Psi_{\Fett p}\rangle&=&p_i|\Psi_{\Fett p}\rangle\label{peigen}
\end{eqnarray}
As is well known it follows from Eqs. (\ref{lie7}) and (\ref{lie9}) that the
eigenvalues $E$ and $p_i$ are not independent, in fact one has
\begin{equation}
E=\sqrt{m^2+{\fett p}^2}\label{energie}
\end{equation}
Here $m$ is the physical rest mass defined in a system where $\fett p=0$. 
Thus with $|\Psi_{\Fett 0}\rangle\equiv|\Psi_{\Fett p{\scriptscriptstyle=}\Fett 0}\rangle$ 
one has
\begin{eqnarray}
H|\Psi_{\Fett 0}\rangle&=&m|\Psi_{\Fett 0}\rangle\\
P_i|\Psi_{\Fett 0}\rangle&=&0
\end{eqnarray}
Now we would like to investigate a realisation in the following form.
We look at a system of two interacting scalar fields $\Psi$ and
$\Phi$ representing nucleons and mesons, 
$\Psi$ being charged and $\Phi$ being a real field. The
interaction we look at is given by
\begin{equation}
{\cal L}_I(x)=g\Psi^\dagger(x)\Psi(x)\Phi(x)\label{lagrange}
\end{equation}
where
\begin{eqnarray}
\Psi(x)&=&{1\over\sqrt{2\pi}^3}\int \hskip-4pt d^3  q{1\over\sqrt{2E_{\Fett q}}}
\left(a_{\Fett q} {\rm e}^{-i qx}+b_{\Fett q}^\dagger{\rm e}^{i
qx}\right)\label{fieldpsi}\\
\Phi(x)&=&{1\over\sqrt{2\pi}^3}\int \hskip-4pt d^3  k{1\over\sqrt{2\omega_{\Fett k}}}
\left(c_{\Fett k} {\rm e}^{-i kx}+c_{\Fett k}^\dagger{\rm e}^{i
kx}\right)\label{fieldphi}\\
E_{\Fett q}&\equiv&\sqrt{m_0^2+\fett q^2}\label{ep}\\
\omega_{\Fett k}&\equiv&\sqrt{\mu_0^2+\fett k^2}\label{omegak}
\end{eqnarray}
Here $m_0$ and $\mu_0$ are the bare masses of nucleons and mesons,
$a^\dagger_{\Fett q}$ creates a nucleon while $b^\dagger_{\Fett q}$
creates an anti-nucleon and the $c_{\Fett k}$ refer to the mesons.

From Eq. (\ref{lagrange}) and with the help of the usual Noether
current arguments one can derive  
expressions for the ten generators
of Poincar\'e transformations in terms of the field operators given in Eqs.
(\ref{fieldpsi}) and (\ref{fieldphi}) and performing space integrals
in terms of the particle creation and
annihilation operators.
In other words we want 
to use the instant form proposed by Dirac \cite{dirac}. 
Due to the usual equal time commutation relations of the fields
(\ref{fieldpsi}) and (\ref{fieldphi}) these generators fulfill the
Poincar\'e algebra (\ref{lie1})-(\ref{lie9}).

The states $|\Psi_{\Fett p}\rangle$ in Eqs. (\ref{heigen}) and 
(\ref{peigen}) are now the physical eigenstates to the field theoretical
model Hamiltonian and total momentum operator. Clearly $H$ will depend
now on the bare masses $m_0$ and $\mu_0$. To calculate the physical mass
$m$ of one nucleon in terms of the bare masses $m_0$, $\mu_0$, and 
the bare coupling constant $g$ 
we want to apply the Okubo transformation \cite{okubo}. 

We regard the field theoretical ten generators to be presented as matrices in
the Fock space:
\begin{equation}
G=G\sum\limits_F|F\rangle\langle F|=\sum\limits_F|F\rangle\langle F|G
\label{matann}
\end{equation}
where $\{|F\rangle\}$ is a complete set of Fock space states and $G$
is any of the ten generators of the Poincar\'e group. Corresponding to
Eq. (\ref{matann}) we introduce Fock space matrix representations of the
generators which we call again $G$. As was shown in \cite{krgl} those
matrices fulfill the set (\ref{lie1}) - (\ref{lie9}). This matrix
representation of the algebra is our starting point.

As was proposed in \cite{glmu} and demonstrated in \cite{krgl} 
the Okubo transformation carried through with one and the
same unitary matrix $\cal U$ block diagonalises the matrices of
all ten generators with
respect to two subspaces of the full Fock space. As a consequence the Lie
algebra of the Poincar\'e group given in Eqs. (\ref{lie1}) -
(\ref{lie9}) is still valid on both of the subspaces separately and we
want to call the projections of the transformed generators on either
of the subspaces "effective generators".

At this point we make the choice that one
of the two sub spaces consists of one nucleon only. Accordingly we define the
projection operators
\begin{equation}
\eta\equiv\int \hskip-4pt d^3  p\ a_{\Fett p}^\dagger|0\rangle\langle0|a_{\Fett p}
\label{eta}
\end{equation}
and
\begin{equation}
\Lambda\equiv 1-\eta
\end{equation}
In accordance to the definition of $\eta$ in Eq. (\ref{eta}) we 
think of $\Lambda$ as being spanned by the set of eigenstates of the
three generators
$P_i$ which are not lying in the $\eta$ space.

Following Eq. (\ref{matann}) we write
\begin{equation}
G=G(\eta+\Lambda)=(\eta+\Lambda)G\label{matgen}
\end{equation}

Then under the action of the Okubo transformation we get
for the four generators $G\equiv H, K_i$
\begin{equation}
G{\buildrel{\cal U}\over\longrightarrow} {\cal U}G{\cal
U}^\dagger\equiv G'\label{genint}
\end{equation}
where $G'$ is now block diagonal:
\begin{equation}
G'=\eta G'\eta+\Lambda G'\Lambda
\end{equation}
Since we are working in the instant form, the generators $P_i$ and $J_i$ are
already diagonal in the Fock space and remain unchanged under $\cal U$
\cite{glmu}.
\begin{eqnarray}
G&{\buildrel{\cal U}\over\longrightarrow}& G\label{genfree}\\
G&=&P_i, J_i
\end{eqnarray}
Now we transform the matrix eigenvalue equations 
\begin{eqnarray}
\langle F|H(\eta+\Lambda)|\Psi_{\Fett p}\rangle&=&E\langle
F|\Psi_{\Fett p}\rangle\\
\langle F|P_i(\eta+\Lambda)|\Psi_{\Fett p}\rangle&=&p_i\langle
F|\Psi_{\Fett p}\rangle
\end{eqnarray}
and get
\begin{eqnarray}
\langle F|H'(\eta+\Lambda)|\Psi'_{\Fett p}\rangle
&=&E\langle F|\Psi'_{\Fett p}\rangle\label{energieeff}\\
\langle F|P_i(\eta+\Lambda)|\Psi'_{\Fett p}\rangle
&=&p_i\langle F|\Psi'_{\Fett p}\rangle\label{impulseff}
\end{eqnarray}
with the transformed eigen state
\begin{equation}
\langle F|\Psi'_{\Fett p}\rangle\equiv\langle F|
{\cal U}(\eta+\Lambda)|\Psi_{\Fett p}\rangle
\end{equation}
Note that the solution vector $|\Psi_{\Fett p}\rangle$ is not
normalisable but the individual components $\langle F|\Psi_{\Fett p}\rangle$
exist, see for instance
\cite{textbook}. 

The new set of generators $G'$ is unitarily equivalent to the original
one so that the eigen value problem presented in Eqs. (\ref{energieeff}) and
(\ref{impulseff}) describes the same physics as the set (\ref{heigen})
and (\ref{peigen}). However in contrast to the original problem (\ref{heigen})
and (\ref{peigen}) we can now focus on solutions $|\phi'_{\Fett p}\rangle$
to (\ref{energieeff}) and
(\ref{impulseff}) which are only lying in the $\eta$ space. These
solutions describe a single freely moving nucleon undergoing
self interactions due to the interaction part in the Lagrangian we
started with given in Eq. (\ref{lagrange}).

Since $\eta$ is the space of one nucleon only and because of
$P_i'=P_i$ we see that $|\phi'_{\Fett p}\rangle$ 
has to be a momentum eigenstate. One has
\begin{equation}
|\phi'_{\Fett p}\rangle=a^\dagger_{\Fett p}|0\rangle\equiv|{\fett p}\rangle
\end{equation}
Due to the coupling of nucleons and mesons occurring in 
the original Hamiltonian the
transformed Hamilton operator $H'$ will carry an interaction:
\begin{equation}
\eta H'\eta=\eta H'\equiv\eta H_0 +\eta H'^{\phantom{i}\mbox{\tiny int}}
\label{hefform}
\end{equation}
where
\begin{equation}
\eta H_0\equiv\int \hskip-4pt d^3  p\ E_{\Fett p}a_{\Fett
p}^\dagger|0\rangle\langle0|a_{\Fett p}\label{h0}
\end{equation}
and $E_{\Fett p}$ being defined in Eq. (\ref{ep}) in terms of the bare
mass.

As we shall show in section \ref{massrenormalisation} $H'$ contains an
infinite constant which is related to the vacuum. Without that
constant the eigenvalue equation (\ref{energieeff}) 
(for an $\langle F|$ which is lying in the $\eta$ space) has to have the form
\begin{equation} 
\left(\eta H_0 +\eta 
H_{\mbox{\tiny nv}}^{'\mbox{\tiny int}}\right)|{\fett p}\rangle
{\buildrel !\over=}\sqrt{m^2+{\fett p}^2}|{\fett p}\rangle\label{start}
\end{equation}
which can be found by inserting Eqs. (\ref{energie}) and
(\ref{hefform})
into Eq. (\ref{energieeff}).
The effective potential carries now the index "nv" (no vacuum) as we
left out the vacuum contribution mentioned above.
Note that the energy eigenvalue $E$ given in Eq. (\ref{energie}) 
is unchanged under the unitary
transformation $\cal U$ and therefore 
the physical
mass enters the square root 
in Eq. (\ref{start}) unlike in the definition of $E_{\Fett q}$
in Eq. (\ref{ep}).

As a consequence of Eq. (\ref{start}) the effective Hamilton operator
without vacuum terms should have the form:
\begin{equation}
\eta H'_{\mbox{\tiny nv}}{\buildrel !\over=}\int \hskip-4pt d^3  p\ \sqrt{m^2+\fett p^2}
a_{\Fett p}^\dagger|0\rangle\langle0|a_{\Fett p}\label{expect}
\end{equation}
Since we are in principle able to calculate 
$\eta H_{\mbox{\tiny nv}}^{'\mbox{\tiny int}}$ 
we can think of Eq. (\ref{start}) as defining $m$ as a function
of the three initial parameters 
$g$, $m_0$, and $\mu_0$. 

It is interesting to see how $\eta H_{\mbox{\tiny nv}}^{'\mbox{\tiny
int}}$ will replace $m_0$ in $H_0$ by the physical mass $m$. This will
be investigated in lowest non trivial order in the next section.

\section{Mass renormalisation}\label{massrenormalisation}
We start with the derivation of $H'$. According to \cite{okubo} and
\cite{krgl} the Okubo transformation (\ref{genint}) leads to
\begin{equation}
\eta H'=(1+A^\dagger A)^{^{-{1\over2}}}(\eta+A^\dagger)H
(\eta+A)(1+A^\dagger A)^{^{-{1\over2}}}\label{eff_form}
\end{equation}
where the operator $A$ has to be of the form
\begin{equation}
A\equiv\Lambda A\eta\label{aform}
\end{equation}
and where the block diagonalisation is guaranteed by the decoupling
equation
\begin{equation}
	\Lambda \biggl([H,A]+H_I-AH_IA\biggr)\eta 
=	0\label{bestH}
\end{equation}
Here we split the original $H$ into
\begin{equation}
H=H_0+H_I
\end{equation}
We solve the non linear equation (\ref{bestH}) perturbatively in
powers of the coupling constant $g$:
\begin{equation}
A=\sum\limits_{\nu=1}^\infty A_\nu\label{reihe1}
\end{equation}
where $A_\nu$ is proportional to $g^\nu$.
In order to calculate $H'$ up to second order in $g$ it is sufficient
to find $A_1$ as follows from Eq. (\ref{eff_form}):
\begin{eqnarray}
\eta H'&=&(1-\halb A^\dagger A)(\eta+A^\dagger)H
(\eta+A)(1-\halb A^\dagger A)+{\cal O}(g^3)\label{schrittheff}\\
&=&(1-\halb A_1^\dagger A_1)
(\eta+A_1^\dagger)H(\eta+A_1)(1-\halb A_1^\dagger A_1)
+{\cal O}(g^3)\label{heff}
\end{eqnarray}
In Eq. (\ref{heff}) we used the property (\ref{aform}) of $A$ 
and also $\eta H_0\Lambda=0=\Lambda H_0\eta$.

Due to Eqs. (\ref{bestH}) and (\ref{reihe1}) one way of finding $A_1$
is to solve the following commutator equation which follows from our
perturbation theoretical treatment of Eq. (\ref{bestH}):
\begin{equation}
{[}H_0,A_1]=-\Lambda H_I\eta\label{bestH1}
\end{equation}
So we need to know $H_I$. In the instant form the
interaction part of the Hamilton operator according to our model given 
in Eq. (\ref{lagrange}) reads:
\begin{eqnarray}
H_I=-{g\over\sqrt {2\pi}^3}\int \hskip-4pt d^3  p\  \hskip-4pt d^3  q\  \hskip-4pt d^3  k\ &&{1\over\sqrt{8E_{\Fett
p}E_{\Fett q}\omega_{\Fett k}}}\nonumber\\
&&\times\biggl[a_{\Fett p}^\dagger a_{\Fett q}c_{\Fett k}^\dagger\delta^3(\fett
p-\fett q+\fett k)
+a_{\Fett p}^\dagger b_{\Fett q}^\dagger c_{\Fett k}^\dagger 
\delta^3(\fett p+\fett q+\fett k)\nonumber\\
&&+a_{\Fett p}^\dagger a_{\Fett q}c_{\Fett k}\delta^3(\fett
p-\fett q-\fett k)
+a_{\Fett p}^\dagger b_{\Fett q}^\dagger c_{\Fett k} 
\delta^3(\fett p+\fett q-\fett k)\nonumber\\
&&+b_{\Fett p}^\dagger b_{\Fett q}c_{\Fett k}^\dagger\delta^3(\fett
p-\fett q+\fett k)
+b_{\Fett p}a_{\Fett q} c_{\Fett k}^\dagger 
\delta^3(-\fett p-\fett q+\fett k)\nonumber\\
&&+b_{\Fett p}^\dagger b_{\Fett q}c_{\Fett k}\delta^3(\fett
p-\fett q-\fett k)
+b_{\Fett p} a_{\Fett q} c_{\Fett k} 
\delta^3(-\fett p-\fett q-\fett k)\biggr]\label{hint}
\end{eqnarray}
and one finds 
\begin{eqnarray}
A_1&=&{g\over\sqrt{2\pi}^3}\int \hskip-4pt d^3  p\  \hskip-4pt d^3  q\  \hskip-4pt d^3  r\  \hskip-4pt d^3  k\
a^\dagger_{\Fett p}a^\dagger_{\Fett r}b^\dagger_{\Fett q}
c^\dagger_{\Fett k}|0\rangle\langle0|a_{\Fett r}
{\delta^3(\fett p+\fett q+\fett k)\over
\sqrt{8E_{\Fett p}E_{\Fett q}\omega_{\Fett k}}
(E_{\Fett p}+E_{\Fett q}+\omega_{\Fett k})}\nonumber\\
&&+{g\over\sqrt{2\pi}^3}\int \hskip-4pt d^3  p\  \hskip-4pt d^3  q\  \hskip-4pt d^3  k\ a_{\Fett p}^\dagger
c_{\Fett k}^\dagger|0\rangle\langle0|a_{\Fett q}{\delta^3(\fett
p-\fett q+\fett k)\over\sqrt{8E_{\Fett p}E_{\Fett q}\omega_{\Fett
k}}(E_{\Fett p}-E_{\Fett q}+\omega_{\Fett k})}\label{a1}
\end{eqnarray}
For the interaction (\ref{hint}) and for
$\eta$ projecting on the space of one nucleon 
there is $\eta H_I\eta=0$ such that Eq. (\ref{heff})
simplifies to give:
\begin{eqnarray}
\eta H'&=&\eta H_0\eta+A_1^\dagger H_I+H_IA_1+A_1^\dagger H_0A_1
-\halb A_1^\dagger A_1H_0-\halb H_0A_1^\dagger A_1\\
&=&\eta H_0+\eta\left(\halb H_IA_1+\halb A_1^\dagger
H_I\right)\eta\label{heff1}
\end{eqnarray}
In the last step (\ref{heff1}) we used Eq. (\ref{bestH1}) and $\eta
H_0=\eta H_0\eta$ which follows from $H_0$ being a free operator. Using
Eqs. (\ref{hint}) and (\ref{a1}) we find
\begin{eqnarray}
\eta H_IA_1\eta=\! -{g^2\over8(2\pi)^3}&&\biggl\{\!\int \hskip-4pt d^3  p\  \hskip-4pt d^3  q\ 
a^\dagger_{\Fett p}|0\rangle\langle0|a_{\Fett p}
{1\over E_{\Fett p}E_{\Fett q}\omega_{\Fett p{\scriptscriptstyle +}\Fett q}}
\Bigl(
{1\over\omega_{\Fett p{\scriptscriptstyle +}\Fett q}+E_{\Fett p}+E_{\Fett q}}
+{1\over\omega_{\Fett p{\scriptscriptstyle +}\Fett q}-E_{\Fett
p}+E_{\Fett q}}
\Bigr)\nonumber\\
&&+\eta\int \hskip-4pt d^3  q\  \hskip-4pt d^3  p\  \hskip-4pt d^3  k\ {\delta^3(\fett p+\fett q+\fett k)
\delta^3(\fett p+\fett q+\fett k)\over E_{\Fett p}E_{\Fett
q}\omega_{\Fett k}}{1\over E_{\Fett p}+E_{\Fett q}+\omega_{\Fett
k}}\biggr\}
\nonumber\\
=-{g^2\over8(2\pi)^3}&&\int \hskip-4pt d^3  p\  \hskip-4pt d^3  q\ 
a^\dagger_{\Fett p}|0\rangle\langle0|a_{\Fett p}
{1\over E_{\Fett p}E_{\Fett q}\omega_{\Fett p{\scriptscriptstyle +}\Fett q}}
\Bigl(
{1\over\omega_{\Fett p{\scriptscriptstyle +}\Fett q}+E_{\Fett p}+E_{\Fett q}}
+{1\over\omega_{\Fett p{\scriptscriptstyle +}\Fett q}-E_{\Fett
p}+E_{\Fett q}}
\Bigr)\nonumber\\
&&+\eta\mbox{\it VAC}\label{ha}
\end{eqnarray}
The second term in Eq. (\ref{ha}) is an infinite constant
multiplied with the projector $\eta$. Since this constant
does not depend on quantum numbers of the nucleon we called it {\it
VAC} and omit it in what follows. Without this constant {\it
VAC} and with help of Eqs. (\ref{h0}) and (\ref{heff1}) we get
\begin{eqnarray}
\eta H'_{\mbox{\tiny nv}}&=&\int \hskip-4pt d^3  p\ a_{\Fett
p}^\dagger|0\rangle\langle0|a_{\Fett p}\nonumber\\
&&\times
\biggl\{E_{\Fett p}-{g^2\over8E_{\Fett p}(2\pi)^3}\int \hskip-4pt d^3  q\ 
{1\over E_{\Fett q}\omega_{\Fett p{\scriptscriptstyle +}\Fett q}}
\Bigl(
{1\over\omega_{\Fett p{\scriptscriptstyle +}\Fett q}+E_{\Fett p}+E_{\Fett q}}
+{1\over\omega_{\Fett p{\scriptscriptstyle +}\Fett q}-E_{\Fett
p}+E_{\Fett q}}
\Bigr)\biggr\}\nonumber\\
&&+{\cal O}(g^3)\label{hefferg}
\end{eqnarray}
Here we again put an index "nv" to indicate that vacuum terms have been
removed from this operator. We also note that the last denominator
cannot vanish for any choice of $\fett p$ and $\fett q$ since energy
and momentum conservation is impossible on a single
vertex described by our Lagrangian presented in Eq. (\ref{lagrange}).

Our task is to verify that this result is of the form of
Eq. (\ref{expect}) and to give a mathematical expression for the
physical mass $m$ in terms of $m_0$, $\mu_0$, and $g$.
To do so we introduce an expansion of $m$ into powers of the coupling
constant $g$:
\begin{equation}
m=m_0+\Delta m=m_0+\sum\limits_{\nu=1}^\infty m_\nu\label{massreihe}\\
\end{equation}
Similar to our notation in Eq. (\ref{reihe1}) the contributions $m_\nu$ are
proportional to $g^\nu$.

This ansatz ensures that in the limit $g\to0$ the physical and
the bare masses are equal. We note that the terms $m_\nu$ are
independent of momenta $\fett p$, of course.
A Taylor expansion of $\sqrt{m^2+{\fett p}^2}$ around $m=m_0$ gives:
\begin{equation}
\sqrt{m^2+\fett p^2}=E_{\Fett p}+{m_0\over E_{\Fett p}} (m_1+m_2)
+{\fett p^2\over2E_{\Fett p}^3}m_1^2+{\cal
O}(g^3)\label{taylor}
\end{equation}
We insert this expression into Eq. (\ref{expect}) and
find expressions for $m_1$ and $m_2$ by comparing equal powers of $g$
in Eqs. (\ref{expect}) and (\ref{hefferg}). We note that in
Eq. (\ref{hefferg}) there is
no contribution being linear in $g$, hence we get:
\begin{equation}
m_1=0
\end{equation}
We insert this into Eq. (\ref{taylor}) and find
\begin{equation}
\sqrt{m^2+{\fett p}^2}=E_{\Fett p}+{m_0\over E_{\Fett p}}m_2+{\cal
O}(g^3)\label{taylor1}
\end{equation}
Comparing Eq. (\ref{expect}) with Eq. (\ref{hefferg}) again and using
Eq. (\ref{taylor1}) this time we find:
\begin{equation}
m_2=-{g^2\over8m_0(2\pi)^3}\int \hskip-4pt d^3  q\ 
{1\over E_{\Fett q}\omega_{\Fett p{\scriptscriptstyle +}\Fett q}}
\Bigl(
{1\over\omega_{\Fett p{\scriptscriptstyle +}\Fett q}+E_{\Fett p}+E_{\Fett q}}
+{1\over\omega_{\Fett p{\scriptscriptstyle +}\Fett q}-E_{\Fett
p}+E_{\Fett q}}
\Bigr)\label{m2}
\end{equation}
Unfortunately, since the integrand
is clearly dependent on the parameter $\fett p$,
this result seems to violate our crucial assumption that
$m_2$ must not depend on external momenta $\fett p$. We also would like
to see consistency of our result with the well know expression for
the mass correction found by Feynman techniques which is of course
independent of initial momenta. For that purpose we look at the result
obtained from Feynman diagrams and we will show in the following
section that this result can indeed be proved to equal ours given by
Eq. (\ref{m2}) which verifies implicitly that Eq. (\ref{m2}) is not
dependent on $\fett p$. A direct and much harder
way to prove the latter
statement is presented in the Appendix.

Summarizing we have shown that indeed Eq. (\ref{expect}) holds in second 
order in the coupling constant.

\section{comparison to mass renormalisation using Feynman
methods}\label{compfeyn} 
We start with the same field theory as specified by
Eqs. (\ref{lagrange}), (\ref{fieldpsi}), and (\ref{fieldphi}). Also we
note that we often use covariant squares $q^2\equiv q^\mu q_\mu$ due to the
covariant Feynman formalism from now on in contrast to
the three dimensional vector products ${\fett q}^2$ of the last section. In
momentum space the Feynman propagators for the charged "nucleon" and the
uncharged meson are given by:
\begin{eqnarray}
i G(p)&\equiv&{i\over p^2-m_0^2+i\epsilon}\label{ferm}\\
i D(k)&\equiv&{i\over k^2-\mu_0^2+i\epsilon}\label{bos}
\end{eqnarray}
So we get the full propagator up to second order in the coupling
constant as
\begin{eqnarray}
i{\cal G}(p)&=&{i\over p^2-m_0^2+i\epsilon}+
{i\over p^2-m_0^2+i\epsilon}
i\Sigma(p){i\over p^2-m_0^2+i\epsilon}+{\cal O}(g^4)\label{fullprop1}
\\
i\Sigma(p)&\equiv&-{g^2\over(2\pi)^4}\int d^4\, q\ i G(q-p)i
D(q)\label{sigmadef}
\end{eqnarray}
Since $\Sigma(p)$ is a scalar it is actually a function $\Sigma(p^2)$. In a
well known manner \cite{mandl} 
using the requirement that ${\cal G}(p)$ has a pole at the
physical mass $m$ one finds the mass shift given by 
\begin{eqnarray}
m_2^{\mbox{\tiny F}}&=&-{1\over2m_0}\Sigma(m^2)\nonumber\\
&=&-{1\over2m_0}\Sigma(m_0^2)+{\cal O}(g^4)
\label{m2feyn}
\end{eqnarray}
The suffix 'F' denotes that we got this result by using Feynman
diagrams.
By construction it is guaranteed this time that $m_2^{\mbox{\tiny F}}$
is not dependent on initial momenta.

We carry out the $q_0$ integration in Eq. (\ref{sigmadef}) in
order to get a three dimensional form:  
\begin{eqnarray}
m_2^{\mbox{\tiny F}}&=&{i g^2\over2(2\pi)^4m_0}\int \hskip-4pt d^3  q\
\int\limits_{-\infty}^\infty d\, q_0\ {1\over\left(q_0-E_{\Fett p}\right)^2 
-\left(\fett q-\fett p\right)^2-m_0^2+i\epsilon}{1\over q_0^2-\fett
q^2-\mu_0^2+i\epsilon}\nonumber\\
&=&{i g^2\over2(2\pi)^4m_0}\int \hskip-4pt d^3  q\
\int\limits_{-\infty}^\infty d\, q_0\ {1\over\left(q_0-E_{\Fett p}\right)^2 
-E^2_{\Fett q{\scriptscriptstyle -}\Fett p}+i\epsilon}
{1\over q_0^2-\omega^2_{\Fett q}+i\epsilon}
\label{step3}
\end{eqnarray}
Here the terms $E_{\Fett p}$ and $E_{\Fett q{\scriptscriptstyle -}\Fett p}$
occur as we calculate $\Sigma(p^2=m_0^2)$.

A simple integration yields 
\begin{eqnarray}
m_2^{\mbox{\tiny F}}&=&
{i^2g^2\over2m_0(2\pi)^3}\int \hskip-4pt d^3  q \Biggl[
{1\over-2E_{\Fett q{\scriptscriptstyle-}\Fett p}}
\,\,{1\over\left(E_{\Fett p}-E_{\Fett q{\scriptscriptstyle-}\Fett p}
\right)^2-\omega_{\Fett q}^2}\nonumber\\
&&\qquad\qquad\qquad\qquad\qquad\qquad
+{1\over\left(E_{\Fett p}+\omega_{\Fett q}\right)^2
-E^2_{\Fett q{\scriptscriptstyle-}\Fett p}}\,\,
{1\over-2\omega_{\Fett q}}\Biggr]\label{step4}
\end{eqnarray}
For the same reason as stated above the denominators cannot vanish.

We go ahead adding zeros in the numerators:
\begin{eqnarray}
m_2^{\mbox{\tiny F}}&=&
{g^2\over4m_0(2\pi)^3}\int \hskip-4pt d^3  q\ {1\over E_{\Fett
q\scriptscriptstyle-\Fett p}\omega_{\Fett q}}
\Biggl[
\halb{2\omega_{\Fett q}+E_{\Fett q\scriptscriptstyle-\Fett p}-E_{\Fett p}
-E_{\Fett q\scriptscriptstyle-\Fett p}+E_{\Fett p}\over
\left(E_{\Fett p}-E_{\Fett q\scriptscriptstyle-\Fett p}-\omega_{\Fett q}
\right)
\left(E_{\Fett p}-E_{\Fett q\scriptscriptstyle-\Fett p}+\omega_{\Fett q}
\right)}\nonumber\\
&&+\halb{2E_{\Fett q\scriptscriptstyle-\Fett p}-E_{\Fett p}-\omega_{\Fett q}
+E_{\Fett p}+\omega_{\Fett q}\over
\left(E_{\Fett p}+\omega_{\Fett q}-E_{\Fett q\scriptscriptstyle-
\Fett p}\right)
\left(E_{\Fett p}+\omega_{\Fett q}+E_{\Fett q\scriptscriptstyle-
\Fett p}\right)}\Biggr]\label{step6}
\end{eqnarray}
and find
\begin{equation}
m_2^{\mbox{\tiny F}}=
{g^2\over8m_0(2\pi)^3}
\int \hskip-4pt d^3  q\ {1\over E_{\Fett q\scriptscriptstyle-\Fett p}\omega_{\Fett
q}}
\Biggl[
{1\over E_{\Fett p}-E_{\Fett q\scriptscriptstyle-\Fett p}
-\omega_{\Fett q}}
-{1\over E_{\Fett p}+E_{\Fett q\scriptscriptstyle-\Fett p}
+\omega_{\Fett q}}
\Biggr]\label{step7}
\end{equation}
Now we see after the substitution $\fett q\to\fett q+\fett p$
that $m_2^{\mbox{\tiny F}}$ is equal to $m_2$ given in Eq. (\ref{m2}).
Apparently in that form (\ref{step7}) or (\ref{m2}) it is far from obvious
that the integral does not depend on $\fett p$. Nevertheless it can be
shown to be true as presented in the Appendix. Also the steps 
backwards from Eq. (\ref{step7}) to Eq. (\ref{step3}) are not obvious
without knowing the final result. However the three dimensional
formalism we are aiming at is much closer to the standard form of a
non relativistic Schr\"odinger equation used in nuclear physics than the
four dimensional off-the-mass-shell structures. Therefore we think
it is justified to put now efforts into three dimensional Hamiltonian
forms derived from field theory.

\section{renormalisation of the boost operators}\label{renbo}
In section \ref{massrenormalisation} we found that the transformed
Hamilton operator in the $\eta$ space, $\eta H'_{\mbox{\tiny{nv}}}$,
up to additive constants looks like the free Hamilton operator in the,
$\eta$ space $\eta H_0$, after replacing the bare nucleon mass $m_0$ by
the physical nucleon mass $m$. A similar situation
should hold true for the three
transformed boost operators, i.e. we expect that $\eta\subsub
Ki{\mbox{\tiny{nv}}}'$, $i=1,2,3$, are equal to $\eta\subsub K0i$
expressed in terms of $m$ instead of $m_0$. 

One reason for expecting that is the
following: The solutions $|\phi'_{\Fett p}\rangle$ to the 
equations (\ref{energieeff}) and (\ref{impulseff}) which lie in the
$\eta$ space represent a free
single nucleon with the physical mass $m$.
This is because we chose $\eta$ to be the space of a single nucleon.
As a consequence the boost operators $\eta\subsub
Ki{\mbox{\tiny{nv}}}'$corresponding to $\eta
H_{\mbox{\tiny{nv}}}'$ should also be free boost operators generating
boosts of a free particle with mass $m$.

Another reason can be addressed by looking at the Lie algebra of the
Poincar\'e group given in Eqs. (\ref{lie1})-(\ref{lie9}): 
In section \ref{conmassren}
we pointed out that this Lie algebra is still valid on the $\eta$
space on its own. While we noted in section \ref{conmassren} that the
generators $P_i$ and $J_i$ remain unchanged under the action of an
Okubo transformation we proved in section \ref{massrenormalisation}
that the transformed Hamilton operator, up to constants, looks like a
free Hamilton operator where the bare mass $m_0$ is replaced by the
physical mass $m$. As the Lie algebra is fulfilled in the case of free
particles we conclude that the corresponding transformed
boost operators should also look like free boost operators expressed
in terms of $m$ instead of $m_0$.

More precisely: 
While the three free boost operators in the $\eta$ space are given by
\begin{equation}
\eta\subsub K0i={i\over2}\eta\int \hskip-4pt d^3  p\ a_{\Fett p}^\dagger\Bigl(
\sqrt{m_0^2+{\fett p}^2}{\partial\over\partial p_i}
+{\partial\over\partial p_i}\sqrt{m_0^2+{\fett p}^2}\Bigr)
a_{\Fett p}\qquad i=1,2,3
\end{equation}
we expect the transformed full boost operators in the $\eta$ space,
after dropping constants related to the vacuum, to look like
\begin{equation}
\eta\subsub Ki{\mbox{\tiny nv}}'
{\buildrel!\over=}
{i\over2}\eta\int \hskip-4pt d^3  p\ a_{\Fett p}^\dagger\Bigl(\sqrt{m^2+{\fett p}^2}
{\partial\over\partial p_i}+{\partial\over\partial
p_i}\sqrt{m^2+{\fett p}^2}
\Bigr)
a_{\Fett p}\qquad i=1,2,3\label{kexp}
\end{equation}
Like in Eq. (\ref{hefform})
we put an index 'nv' which indicates that we did not take any
constants into account.
We are now going to verify that Eq. (\ref{kexp}) holds true indeed.

According to \cite{okubo} and \cite{krgl} one finds the following expression
for the transformed boost operators in the $\eta$ space after an Okubo
transformation:
\begin{equation}
\eta K_i'
=\eta \subsub K0i+\eta\left(\halb \subsub KIiA_1+\halb A_1^\dagger
\subsub KIi\right)\eta\label{keff1}
\end{equation}
where $A_1$ is given by Eq. (\ref{a1}).
This result corresponds to Eq. (\ref{heff1}).
To go ahead one 
derives an expression for $\subsub KIi$ from 
the interaction part of the Lagrangian given in Eq. (\ref{lagrange}):
\begin{eqnarray}
\subsub KIi=-i{g\over\sqrt {2\pi}^3}\int \hskip-4pt d^3  p\
\hskip-4pt d^3  q\  \hskip-4pt d^3  k\! &&{1\over\sqrt{8E_{\Fett
p}E_{\Fett q}\omega_{\Fett k}}}\nonumber\\
&&\times\biggl[a_{\Fett p}^\dagger a_{\Fett q}c_{\Fett k}^\dagger
{\partial\over\partial k_i}\delta^3(\fett
p-\fett q+\fett k)
+a_{\Fett p}^\dagger b_{\Fett q}^\dagger c_{\Fett k}^\dagger 
{\partial\over\partial k_i}\delta^3(\fett p+\fett q+\fett k)\nonumber\\
&&-a_{\Fett p}^\dagger a_{\Fett q}c_{\Fett k}
{\partial\over\partial k_i}\delta^3(\fett
p-\fett q-\fett k)
-a_{\Fett p}^\dagger b_{\Fett q}^\dagger c_{\Fett k} 
{\partial\over\partial k_i}\delta^3(\fett p+\fett q-\fett k)\nonumber\\
&&+b_{\Fett p}^\dagger b_{\Fett q}c_{\Fett k}^\dagger
{\partial\over\partial k_i}\delta^3(\fett
p-\fett q+\fett k)
+b_{\Fett p}a_{\Fett q} c_{\Fett k}^\dagger 
{\partial\over\partial k_i}\delta^3(-\fett p-\fett q+\fett k)\nonumber\\
&&-b_{\Fett p}^\dagger b_{\Fett q}c_{\Fett k}
{\partial\over\partial k_i}\delta^3(\fett
p-\fett q-\fett k)
-b_{\Fett p} b_{\Fett q} c_{\Fett k} 
{\partial\over\partial k_i}
\delta^3(-\fett p-\fett q-\fett k)\biggr]\label{kint}
\end{eqnarray}
Note the formal similarity to $H_I$ given in Eq. (\ref{hint}).

The expression (\ref{kint}) can be inserted into Eq. (\ref{keff1}).
Again it turns out that $K_i'$ contains an additive infinite
constant being related to the vacuum. After dropping this constant one
arrives at an intermediate result:
\begin{eqnarray}
\subsub Ki{\mbox{\tiny nv}}'
&=&\eta\subsub K0i\nonumber\\
&&-{i g^2\over16E_{\Fett p}(2\pi)^3} 
\int \hskip-4pt d^3  p\  \hskip-4pt d^3  q\ 
{1\over E_{\Fett q}\omega_{\Fett p{\scriptscriptstyle +}\Fett q}}
\Bigl(
{1\over\omega_{\Fett p{\scriptscriptstyle +}\Fett q}+E_{\Fett p}+E_{\Fett q}}
+{1\over\omega_{\Fett p{\scriptscriptstyle +}\Fett q}+E_{\Fett
p}-E_{\Fett q}}
\Bigr)\nonumber\\
&&\times \biggl\{a_{\Fett
p}^\dagger|0\rangle\langle0|{\partial\over\partial p_i} a_{\Fett p}
-\biggl({\partial\over\partial p_i}a_{\Fett
p}^\dagger\biggr)|0\rangle\langle0|a_{\Fett p}\biggr\}
+{\cal O}(g^3)\label{keff2}
\end{eqnarray}
After rewriting $\eta\subsub K0i$ into
\begin{equation}
\eta\subsub K0i=\mbox{\large${i\over2}$}\int \hskip-4pt d^3  p\ E_{\Fett p}
\biggl(a_{\Fett p}^\dagger|0\rangle{\partial\over\partial
p_i}\langle0|a_{\Fett p}-\Bigl({\partial\over\partial p_i}
a_{\Fett p}^\dagger\Bigr)|0\rangle\langle0|a_{\Fett p}\biggr)\label{k0neu}
\end{equation}
we can modify Eq. (\ref{keff2}) to give
\begin{eqnarray}
\eta \subsub Ki{\mbox{\tiny nv}}'&=&\!\mbox{\large${i\over2}$}\!
\int \hskip-4pt d^3  p\ 
\biggl\{E_{\Fett p}-{g^2\over8E_{\Fett p}(2\pi)^3}\int \hskip-4pt d^3  q\ 
{1\over E_{\Fett q}\omega_{\Fett p{\scriptscriptstyle +}\Fett q}}
\Bigl(
{1\over\omega_{\Fett p{\scriptscriptstyle +}\Fett q}+E_{\Fett p}+E_{\Fett q}}
+{1\over\omega_{\Fett p{\scriptscriptstyle +}\Fett q}+E_{\Fett
p}-E_{\Fett q}}
\Bigr)\biggr\}\nonumber\\
&&\times\biggl(a_{\Fett p}^\dagger|0\rangle{\partial\over\partial
p_i}\langle0|a_{\Fett p}-\Bigl({\partial\over\partial p_i}
a_{\Fett p}^\dagger\Bigr)|0\rangle\langle0|a_{\Fett p}\biggr)
+{\cal O}(g^3)\label{kefferg}
\end{eqnarray}
We note that the curly bracket is identical to the one in Eq. (\ref{hefferg})
for $\eta H'_{\mbox{\tiny nv}}$.
As a consequence we see that the integral in $\fett q$ is closely related
to the definition of $m_2$ given in Eq. (\ref{m2}) and we can rewrite
Eq. (\ref{kefferg}) in terms of $m_2$.
\begin{equation}
\eta \subsub Ki{\mbox{\tiny nv}}'=\mbox{\large${i\over2}$}
\int \hskip-4pt d^3  p\ 
\biggl(E_{\Fett p}+{m_0\over E_{\Fett p}}m_2\biggr)
\biggl(a_{\Fett p}^\dagger|0\rangle{\partial\over\partial
p_i}\langle0|a_{\Fett p}-\Bigl({\partial\over\partial p_i}
a_{\Fett p}^\dagger\Bigr)|0\rangle\langle0|a_{\Fett p}\biggr)
+{\cal O}(g^3)\label{kefferg1}
\end{equation}
Now we make use of an intermediate result presented in
Eq. (\ref{taylor1}) and find as the desired result
\begin{equation}
\eta \subsub Ki{\mbox{\tiny nv}}'=\mbox{\large${i\over2}$}
\int \hskip-4pt d^3  p\ 
\biggl(\sqrt{m^2+{\fett p}^2}\biggr)
\biggl(a_{\Fett p}^\dagger|0\rangle{\partial\over\partial
p_i}\langle0|a_{\Fett p}-\Bigl({\partial\over\partial p_i}
a_{\Fett p}^\dagger\Bigr)|0\rangle\langle0|a_{\Fett p}\biggr)
+{\cal O}(g^3)\label{kefferg2}
\end{equation}
which proves that the three transformed boost operators $\eta \subsub
Ki{\mbox{\tiny nv}}'$ look like free boost operators where 
the bare nucleon mass $m_0$ has been replaced by the physical mass
$m$. Then the Poincar\'e algebra is of course fulfilled.

\section{Summary}\label{summary}
Starting from a Lagrangian describing charged scalar "nucleons"
and uncharged mesons which interact via a simple vertex 
expression the ten effective generators for Poincar\'e transformations for 
one nucleon are derived in lowest nontrivial order in the coupling
constant. We used the Okubo transformation
The dynamics of this system of one nucleon is governed
by the generator for time translations, the effective
Hamilton operator. Looking at the simultaneous eigenstates to the
Hamilton and the three momentum operators general arguments lead to
the statement that the eigenvalues with respect to the Hamilton operator 
have to be of the form $E=\sqrt{m^2+\fett{p}^2}$ where $p_i$ are the
eigenvalues to the three components of the momentum operator and $m$
is the total rest mass of the system being described. In our case that
system is the system of one single nucleon undergoing self interactions
due to the interaction part in the Lagrangian proposed above and so
the total rest mass of that system is the total rest mass of a
physical nucleon. 

We used that last statement to derive an expression for that physical
mass in terms of the three initial parameters, the bare masses of the
nucleon and the meson $m_0$ and $\mu_0$ and the coupling constant
$g$. We did this in a perturbation theoretical manner and gave an
expression for $m$ in section \ref{massrenormalisation}
which is correct up to second order in the coupling
constant. After some
calculation we could show that
$m$, at least up to second order, 
is not dependent on the momentum of the particle. This is in
contrast to a first look at the analytical expression for $m$.

In section
\ref{compfeyn} we compared this result to the expression found for $m$
using Feynman techniques and revealed that the two calculations lead to
the same result. The result gained by Feynman techniques
is strictly not dependent on the momentum
of the particle.

By explicit calculation in section \ref{renbo} we showed that
the three effective boost operators can be rewritten in terms of the
physical mass of the nucleon $m$, again we did this in second order in
the coupling constant only. The results are three free effective
boost operators
where the bare mass of the nucleon is just replaced by the physical
mass. 

In the Appendix that question of the '$\fett p$
independence' of the mass posed in section \ref{compfeyn} was answered
without knowledge of the Feynman result by a lengthy but straight
forward calculation.

This completes our discussion of this model of one physical
nucleon stating that up to second order the effective 
generators of Poincar\'e
transformations are equal to the well known free generators of the
Poincar\'e group except that in the Hamilton and the three boost
operators the bare mass of the nucleon is replaced by its physical
mass. 

Clearly the next step will be to investigate two interacting nucleons.
In \cite{glmu} it has been shown that in leading order ${\cal O}(g^2)$
the effective generators fulfill the Poincar\'e algebra. But this does
not yet involve loop integrals and thus renormalizations. They will occur
in the order ${\cal O}(g^4)$. The interesting question will be, whether the
effective generators after renormalization (in the Hamiltonian formalism)
will retain their property formally found in \cite{krgl} to fulfill the
Poincar\'e algebra. This work is under investigation.

\section{Appendix}
In section \ref{compfeyn} we have seen that the second order
mass shift calculated by the
method of unitary transformation in the Hamilton formalism is equal to
the mass shift resulting from a calculation by Feynman methods. Hence
we have shown at the same time that our expression given in
Eq. (\ref{m2}) is not dependent on the initial momentum $\fett p$ which
is in contrast to a first glance at Eq. (\ref{m2}) where the integrand
is a function of $\fett p$.
Now a lengthy calculation which does not make use of results which
have been derived in the framework of a covariant formalism
shows that after the integration this dependence on $\fett p$ is lost.
This will be shown now.
First we notice that actually the integral occurring in Eq. (\ref{m2})
is ultra-violet divergent. We abbreviate and introduce a cutoff $\lambda$:
\begin{equation}
\int\limits_{|\Fett q|{\scriptscriptstyle\le\lambda}}
 \hskip-4pt d^3  q\ F(\fett p,\fett q)\equiv
\int\limits_{|\Fett q|{\scriptscriptstyle\le\lambda}} \hskip-4pt d^3  q\ 
{1\over E_{\Fett q}\omega_{\Fett p{\scriptscriptstyle +}\Fett q}}
\Bigl(
{1\over\omega_{\Fett p{\scriptscriptstyle +}\Fett q}+E_{\Fett p}+E_{\Fett q}}
+{1\over\omega_{\Fett p{\scriptscriptstyle +}\Fett q}-E_{\Fett
p}+E_{\Fett q}}
\Bigr)\label{Cutoff}
\end{equation}
so that in the limit $\lambda\to\infty$ we regain the integral
presented in Eq. (\ref{m2}).
Now we split the convergent (regularized)
integral given in Eq. (\ref{Cutoff}):
\begin{equation}
\int\limits_{|\Fett q|{\scriptscriptstyle\le\lambda}} \hskip-4pt d^3  q\ F(\fett p,\fett q)
=\int\limits_{|\Fett q|{\scriptscriptstyle\le\lambda}} \hskip-4pt d^3  q\
\bigl(F(\fett p,\fett q)-F(0,\fett q)\bigr)
+\int\limits_{|\Fett q|{\scriptscriptstyle\le\lambda}} \hskip-4pt d^3  q\ 
F(0,\fett q)\label{split}
\end{equation}
It can be shown that the first integral is no longer ultra-violet
divergent as $\lambda\to\infty$:
\begin{equation}
\lim_{\lambda\to\infty}
\int\limits_{|\Fett q|{\scriptscriptstyle\le\lambda}} \hskip-4pt d^3  q\
\bigl(F(\fett p,\fett q)-F(0,\fett q)\bigr)<\infty\label{absch}
\end{equation}
We see that 
the "$\fett p$ dependence" of $m_2$ is now
accessible to investigation since its origin is the first, well
defined expression in Eq. (\ref{split}). Consequently, to investigate
the $\fett p$ dependence of $\int \hskip-4pt d^3  q\ F(\fett q,\fett p)$ we now draw our
attention to the first term.

We can simplify the integrals analytically:
\begin{eqnarray}
\int\limits_{|\Fett q|{\scriptscriptstyle\le\lambda}}\!\!
 \hskip-4pt d^3  q\ F(0,\fett q)&=&8\pi\int\limits_0^\lambda\!\! d\, q\
{q^2\over\sqrt{\mu_0^2+q^2}\sqrt{m_0^2+q^2}}
{\sqrt{\mu_0^2+q^2}+\sqrt{m_0^2+q^2}\over 
(\sqrt{\mu_0^2+q^2}+\sqrt{m_0^2+q^2})^2-m_0^2}\label{f0}\\
\int\limits_{|\Fett q|{\scriptscriptstyle\le\lambda}}\!\!
 \hskip-4pt d^3  q\ F(\fett p,\fett q)&=&
2\pi\!\!\int\limits_0^\lambda\!\! d\, q\
{q\over\sqrt{m_0^2+q^2}p}\ln
{\left(\sqrt{m_0^2+p^2}+\sqrt{\mu_0^2+(p+q)^2}\right)^2-m_0^2-p^2\over
 \left(\sqrt{m_0^2+p^2}+\sqrt{\mu_0^2+(p-q)^2}\right)^2-m_0^2-p^2}
\label{fp}
\end{eqnarray}
where
\begin{eqnarray}
p&\equiv&|\fett p|\\
q&\equiv&|\fett q|
\end{eqnarray}
We introduce dimensionless quantities
\begin{eqnarray}
x&\equiv&{q\over m_0}\label{def00}\\
\alpha&\equiv&{p\over m_0}\label{def01}\\
\beta_0&\equiv&{\mu_0\over m_0}\label{def02}\\
\Lambda&\equiv&{\lambda\over p}\label{def03}
\end{eqnarray}
and define
\begin{eqnarray}
{\cal I}_\alpha^\Lambda(\beta)
\equiv2\pi\int\limits_0^\Lambda d\, x&&\Biggl\{
4{x^2\over\sqrt{1+x^2}\sqrt{\beta^2+x^2}}
{\sqrt{1+x^2}+\sqrt{\beta^2+x^2}\over
\left(\sqrt{1+x^2}+\sqrt{\beta^2+x^2}\right)^2-1}\nonumber\\
&&-{1\over\alpha}{x\over\sqrt{1+x^2}}\ln
{\left(\sqrt{1+x^2}+\sqrt{\beta^2+(x+\alpha)^2}\right)^2-1-\alpha^2\over
\left(\sqrt{1+x^2}+\sqrt{\beta^2+(x-\alpha)^2}-1-\alpha^2\right)^2}\Biggr\}
\label{term1}
\end{eqnarray}
Then find
\begin{equation}
\lim_{\Lambda\to\infty}{\cal I}_\alpha^\Lambda(\beta_0)=
\int  \hskip-4pt d^3  q\ \left(F(0,\fett q)-F(\fett p,\fett q)\right)\label{ibeta0}
\end{equation}
For any value of $\alpha$ and $\Lambda$
we want to think of ${\cal I}_\alpha^\Lambda(\beta)$ as being an 
analytical function dependent on the complex
parameter $\beta$. Due to Eq. (\ref{absch}) this also includes the
case $\Lambda=\infty$. From Eq. (\ref{ibeta0}) we see  
that ${\cal I}_\alpha^\Lambda(\beta)$ becomes a physical
meaning at the point $\beta_0$ defined by
Eq. (\ref{def02}) and in the limit $\Lambda=\infty$.  

For the case of $\lambda<\infty$ 
we want to modify the second term of Eq. (\ref{ibeta0}):
\begin{eqnarray}
\lefteqn{-\int\limits_{|\Fett q|{\scriptscriptstyle\le}\lambda} \hskip-4pt d^3  q\ F(\fett p,\fett q)}
\nonumber\\
&=&{-2\pi\over\alpha}\int\limits_0^\Lambda d\,  x
{x\over\sqrt{1+x^2}}\ln
{\left(\sqrt{1+x^2}+\sqrt{\beta^2+(x+\alpha)^2}\right)^2-1-\alpha^2\over
\left(\sqrt{1+x^2}+\sqrt{\beta^2+(x-\alpha)^2}\right)^2-1-\alpha^2}
\nonumber\\
&=&{-2\pi\over\alpha}\int
\limits_{-\Lambda}^\Lambda d\,  x
{x\over\sqrt{1+x^2}}\ln
\left(\left(\sqrt{1+x^2}+\sqrt{\beta^2+(x+\alpha)^2}\right)^2-1-\alpha^2
\right)\nonumber\\
&=&{-2\pi\over\alpha}\sqrt{1+x^2}\ln
\left\{\left(\sqrt{1+x^2}+\sqrt{\beta^2+(x+\alpha)^2}\right)^2-1-\alpha^2
\right\}\Biggl|_{-\Lambda}^\Lambda\\
&+&{2\pi\over\alpha}\int
\limits_{-\Lambda}^\Lambda d\,  x\sqrt{1+x^2}
{2\left(\sqrt{1+x^2}+\sqrt{\beta^2+(x+\alpha)^2}\right)
\left({x\over\sqrt{1+x^2}}+{x+\alpha\over\sqrt{\beta^2+(x+\alpha)^2}}\right)
\over\left(\sqrt{1+x^2}+\sqrt{\beta^2+(x+\alpha)^2}\right)^2-1-\alpha^2}
\nonumber
\end{eqnarray}
Here the first of the two terms is equal to $-4\pi$ as $\Lambda$ goes to
infinity. So we reformulate Eq. (\ref{term1}) as:
\begin{eqnarray}
\lim_{\Lambda\to\infty}{\cal I}_\alpha^\Lambda(\beta)\nonumber\\
=-4\pi&&+2\pi\lim_{\Lambda\to\infty}
\int\limits_{-\Lambda}^\Lambda d\,  x\Biggl\{
{2x^2\over\sqrt{1+x^2}\sqrt{\beta^2+x^2}}
{\sqrt{1+x^2}+\sqrt{\beta^2+x^2}\over
\left(\sqrt{1+x^2}+\sqrt{\beta^2+x^2}\right)^2-1}\nonumber\\
&&+{2\sqrt{1+x^2}\over\alpha}
{\left(\sqrt{1+x^2}+\sqrt{\beta^2+(x+\alpha)^2}\right)
\left({x\over\sqrt{1+x^2}}+{x+\alpha\over\sqrt{\beta^2+(x+\alpha)^2}}\right)
\over\left(\sqrt{1+x^2}+\sqrt{\beta^2+(x+\alpha)^2}\right)^2-1-\alpha^2}
\Biggr\}
\end{eqnarray}
It can be shown that no singularities are lying on the path of integration. To
solve this integral we introduce the following substitution:
\begin{equation}
\sqrt{x^2+1}\equiv z(x-i)
\label{subst}
\end{equation}
This implies that the path of integration is shifted into the upper
half of the complex
$z$-plane going from $-1$ to $+1$ on a half
circle. After the substitution the integrand is a rational function
which can be integrated analytically. We also want to note 
that it is important
to treat the limits correctly. The
result is:
\begin{eqnarray}
&&\lim_{\Lambda\to\infty}{\cal I}_\alpha^\Lambda(\beta)\nonumber\\
&&=-4\pi+2\pi\lim_{\Lambda\to\infty}\Biggl[
-\beta^4\int\limits_
{\sqrt{{-\Lambda+\beta i\over-\Lambda-\beta i}}}^
{\sqrt{{\Lambda+\beta i\over\Lambda-\beta i}}}
 d\,  z{1\over z^2-1}
{i^2{(z^2+1)^2\over(z^2-1)^2}
\over
\left(i{z^2+1\over z^2-1}\beta-\zeta\right)
\left(i{z^2+1\over z^2-1}\beta+\zeta\right)
}\nonumber\\
&&+\int\limits_
{\sqrt{-\Lambda+i\over-\Lambda-i}}^
{\sqrt{\Lambda+i\over\Lambda-i}}
 d\,  z{1\over z^2-1}
{i^2{(z^2+1)^2\over(z^2-1)^2}(\beta^2-2)
\over
\left(i{z^2+1\over z^2-1}-\zeta\right)
\left(i{z^2+1\over z^2-1}+\zeta\right)
}
\nonumber\\
-&&{4\over\alpha}\int\limits_
{\sqrt{-\Lambda+i\over-\Lambda-i}}^
{\sqrt{\Lambda+i\over\Lambda-i}} d\,  z
{z^2\over\left(z^2-1\right)^3}{-2i{z^2+1\over z^2-1}+\alpha(\beta^2-2)\over
\left(i{z^2+1\over z^2-1}-\eta+\alpha-\xi\right)
\left(i{z^2+1\over z^2-1}-\eta+\alpha+\xi\right)}\nonumber\\
-&&{4\over\alpha}\int\limits_
{\sqrt{-\Lambda+\alpha+\beta i\over-\Lambda+\alpha-\beta i}}^
{\sqrt{\Lambda+\alpha+\beta i\over\Lambda+\alpha-\beta i}} d\,  z
{z^2\over\left(z^2-1\right)^3}
{1\over i\left({z^2+1\over z^2-1}-1\right)}
{1\over i\left({z^2+1\over z^2-1}+1\right)}
{1\over\left(i\beta{z^2+1\over z^2-1}-\eta-\xi\right)}
{1\over\left(i\beta{z^2+1\over z^2-1}-\eta+\xi\right)}
\nonumber\\
&&\phantom{AAAAAAAA}
\times\Biggl(2\alpha\beta^2+2\alpha^3\beta^2-\alpha\beta^4
+(-3\beta^3-5\alpha^2\beta^3+\beta^5)i{z^2+1\over z^2-1}
\nonumber\\
&&\phantom{AAAAAAAAAAAAA}
+3\alpha\beta^4i^2{(z^2+1)^2\over (z^2-1)^2}
-2\beta^3i^3{(z^2+1)^3\over (z^2-1)^3}\Biggr)
\Biggr]\label{integral}
\end{eqnarray}
where we used:
\begin{eqnarray}
\xi&\equiv&{\beta\over2}\sqrt{(\alpha^2+1)(\beta^2-4)}\label{def1}\\
\eta&\equiv&{\alpha\beta^2\over2}\label{def2}\\
\zeta&\equiv&{\beta\over2}\sqrt{\beta^2-4}\label{def3}
\end{eqnarray}
For $\beta\ge2$ we see that $\xi$ and $\zeta$ remain real and we want
to concentrate on that case first.
The integral Eq. (\ref{integral}) can be carried out analytically and
we find for $\beta\ge2$:
\begin{eqnarray}
\lefteqn{\lim_{\Lambda\to\infty}{\cal I}_\alpha^\Lambda(\beta)}\nonumber\\
&=&-2\pi\Biggl[
-{1\over\alpha}\sqrt{1+(\eta-\alpha+\xi)^2}
\ln\left(\sqrt{1+(\eta-\alpha+\xi)^2}-\eta+\alpha-\xi\right)\nonumber\\
&&-{1\over\alpha}\sqrt{1+(\eta-\alpha-\xi)^2}
\ln\left(\sqrt{1+(\eta-\alpha-\xi)^2}-\eta+\alpha+\xi\right)\nonumber\\
&&+{1\over4\alpha}
{2\alpha^2\beta^2(-3+\beta^2)+(-2+\beta^2)\beta^2+4\alpha(\beta^2-1)\xi
\over\sqrt{\beta^2+(\eta+\xi)^2}}
\ln{\sqrt{\beta^2+(\eta+\xi)^2}-\eta-\xi\over\beta}\nonumber\\
&&+{1\over4\alpha}
{2\alpha^2\beta^2(-3+\beta^2)+(-2+\beta^2)\beta^2-4\alpha(\beta^2-1)\xi
\over\sqrt{\beta^2+(\eta-\xi)^2}}
\ln{\sqrt{\beta^2+(\eta-\xi)^2}-\eta+\xi\over\beta}\nonumber\\
&&-{2\beta}\sqrt{{\beta^2\over4}-1}
\ln\left(\sqrt{{\beta^2\over4}-1}+{\beta\over2}\right)
\Biggr]\label{stamm1}
\end{eqnarray}
To simplify this expression we employ tho following substitution \cite{kamada}:
\begin{eqnarray}
\beta&\equiv&{u^2+1\over u}\label{defa1}\\
\alpha&\equiv&{2nm\over n^2-m^2}\label{defa2}
\end{eqnarray}
In terms of these new variables we get rid of all the square roots in
Eq. (\ref{stamm1}) and also the arguments of the logarithms simplify
greatly:
\begin{eqnarray}
\lim_{\Lambda\to\infty}{\cal I}_\alpha^\Lambda(\beta)=-2\pi\Biggl[
&&-{(m-n)^2+(m+n)^2u^4\over4mnu^2}\left(\ln{-m+n\over m+n}-2\ln u\right)
\nonumber\\
&&-{(m+n)^2+(m-n)^2u^4\over4mnu^2}\left(\ln{-m+n\over m+n}+2\ln u\right)
\nonumber\\
&&+{(m-n)^2+(m+n)^2u^4\over4mnu^2}\left(\ln{-m+n\over m+n}-\ln u\right)
\nonumber\\
&&+{(m+n)^2+(m-n)^2u^4\over4mnu^2}\left(\ln{-m+n\over m+n}+\ln u\right)
\nonumber\\
&&+{1-u^4\over u^2}\ln u\Biggr]
\label{stamm2}
\end{eqnarray}
One easily checks by comparing the coefficients of equal logarithms 
that we get the final result
\begin{eqnarray}
\lim_{\Lambda\to\infty}{\cal I}_\alpha^\Lambda(\beta)&=&0\label{stamm3}\\
\beta&\ge&2\label{condition}
\end{eqnarray}
Finally, since $\lim_{\Lambda\to\infty}{\cal I}_\alpha^\Lambda(\beta)$
is an analytical function
which is zero over some finite interval on the real axis, we conclude
that 
\begin{equation}
\lim_{\Lambda\to\infty}{\cal I}_\alpha^\Lambda(\beta)\equiv0\label{stamm4}
\end{equation}
everywhere on the complex $\beta$ plane and thus also for
$\beta=\beta_0$. 
Because of Eq. (\ref{ibeta0})
we can then rewrite Eq. (\ref{split}) as
\begin{equation}
\int \hskip-4pt d^3  q\ F(\fett p,\fett q)=\int \hskip-4pt d^3  q\ F(0,\fett q)\label{pindep}
\end{equation}
demonstrating clearly the independence of $\fett p$.
As a consequence we can 
give an expression for $m_2$ which is now, in contrast to
Eq. (\ref{m2}), independent of the initial momentum $\fett p$:
\begin{equation}
m_2=-{g^2\over8m_0(2\pi)^3}\int \hskip-4pt d^3  q\ {1\over E_{\Fett q}\omega_{\Fett
q}}\left({1\over\omega_{\Fett q}+m_0+E_{\Fett q}}+
{1\over\omega_{\Fett q}-m_0+E_{\Fett q}}\right)
\label{m2final}
\end{equation}
We arrive then indeed at Eq. (\ref{expect}):
\begin{equation}
\eta H'_{\mbox{\tiny nv}}=\int \hskip-4pt d^3  p\ \sqrt{m^2+\fett p^2}
a_{\Fett p}^\dagger|0\rangle\langle0|a_{\Fett p}+{\cal O}(g^3)
\label{expect1}
\end{equation}
with
\begin{equation}
m=m_0-{g^2\over8m_0(2\pi)^3}\int \hskip-4pt d^3  q\ {1\over E_{\Fett q}\omega_{\Fett
q}}\left({1\over\omega_{\Fett q}+m_0+E_{\Fett q}}+
{1\over\omega_{\Fett q}-m_0+E_{\Fett q}}\right)+{\cal O}(g^3)
\end{equation}
We want to point out that this result has been derived in the
framework 
of a Hamilton formalism and did not use the connection to the standard 
techniques.

\end{document}